\documentstyle[amssymb,12pt,manuscript,aps,epsfig]{revtex}

\begin{document}
\title{Multivariate distribution of returns in financial time series}
\author{E. Alessio$^{\dagger )}$, V. Frappietro$^{\dagger )}$, M. I. Krivoruchenko$%
^{\dagger ,*)}$, L. J. Streckert$^{\dagger )}$ \\
$^{\dagger )}$Metronome-Ricerca sui Mercati Finanziari, C. so Vittorio
Emanuele 84\\
10121 Torino, Italy\\
$^{*)}$Institute for Theoretical and Experimental Physics, B.
Cheremushkinskaya 25\\
117259 Moscow, Russia}
\maketitle

\begin{abstract}
Multivariate probability density functions of returns are constructed in order to model the empirical behavior of returns in a financial time series.  They describe the well-established deviations from the Gaussian random walk, such as an approximate scaling and heavy tails of the return distributions, long-ranged volatility-volatility correlations (volatility clustering) and return-volatility correlations (leverage effect).  Free parameters of the model are fixed over the long term by fitting 100+ years of daily prices of the Dow Jones 30 Industrial Average.  The multivariate probability density functions which we have constructed can be used for pricing derivative securities and risk management.

\ \ $\;\;\;$\newline
{}


{\bf PACS:} \hspace{0pt}\ \ 89.65.Gh, 89.75.Da, 02.50.Ng, 02.50.Sk 
\end{abstract}
\newpage{}

The methods developed in studying complex physical systems have been
successfully applied throughout decades to analize financial data \cite
{bachelier,levy,mandelbrot} and they continue to attract gradual interest \cite
{hull,para,ales,bouchaud,pagan,fama,mantegna,manst}. The field of research
connected to modeling financial markets and development of statistically
based real-time decision systems has recently been named Econophysics. The
properties of certain observables of different financial markets appear to
be quite similar, indicating the existence of universal driving mechanisms
behind market evolution. The understanding of the dynamics of complex
financial systems has an influence on the decisions of market participants and
on the market behavior and represents a scientific challenge for research.
The methods of Econophysics have made previous progress connected to
the occurrence of quantitatively based financial management firms and the
development of high performance online systems which
use methods of statistical inference from real-time and
historical prices as well as other economic information.

The behavior of a financial time series can be described by a generalized Wiener
process. In terms of a drift rate $\mu $, the evolution of a market index
value or a stock price $S(t)$ is governed by equation \cite{hull}: 
\begin{equation}
\frac{dS(t)}{S(t)}=\mu dt+d\xi (t).  \label{dS/S}
\end{equation}
The value $d\xi (t)$ is a noise added to the path followed by $S(t)$ with
the expectation value and variance of
\begin{eqnarray}
{\rm E}[d\xi (t)] &=&0,  \label{noise1} \\
{\rm Var}[d\xi (t)] &=&\sigma (t)^{2}dt.  \label{noise2}
\end{eqnarray}
The volatility $\sigma (t)$ represents a generic measure of the magnitude of
market fluctuations. It quantifies risk and enters as an input to option
pricing models. We consider a discrete random walk and set $dt\equiv
t_{i}-t_{i-1},$ $S_{i}\equiv S(t_{i}),$ $\xi _{i}\equiv d\xi (t_{i}),\,$and $%
\sigma _{i}\equiv \sigma (t_{i})$.

The random walk model proposed by Bachelier in the year 1900 \cite{bachelier} is
equivalent to the Gaussian multivariate probability density function (PDF)
of the increments $\xi _{i}$: 
\begin{equation}
G_{n}(\xi _{i})=\frac{1}{(2\pi )^{n/2}}\exp (-\frac{1}{2}\sum_{i=1}^{n}%
\sigma _{i}^{-2}\xi _{i}^{2})\prod_{i=1}^{n}\frac{1}{\sigma _{i}}.
\label{gauss}
\end{equation}
The statistically significant correlations of the increments $\xi _{i}$ are
absent from a time frame greater than $t_{i}-t_{i-1}=$ $20$ Min \cite{para}. The absence of
the correlations,
\begin{equation}
{\rm Corr}[\xi _{i},\xi _{j}]=\delta _{ij},  \label{zero}
\end{equation}
has been widely documented (see e.g. Ref. \cite{manst}) and is often cited as a
support for the efficient market hypothesis \cite{fama}. The model (\ref
{gauss}) constitutes a solid zero-order approximation to the empirical
distributions of the increments $\xi _{i}$. The multivariate PDFs
constructed in this work are extensions of the Gaussian PDFs, aimed to
model the well-established deviations in the behavior of financial time
series from the Gaussian random walk.

The L\'{e}vy stable truncated univariate PDFs \cite{levy,mantegna,manst} are
known to provide, for a financial time series, (i) an approximate scaling
invariance of the univariate PDFs with a slow convergence to the Gaussian
behavior and (ii) the existence of heavy tails. We propose the
multivariate Student PDFs, 
\begin{equation}
S_{n}^{\alpha }(\xi _{i})=\frac{\Gamma (\frac{\alpha +n}{2})}{(\alpha \pi
)^{n/2}\Gamma (\frac{\alpha }{2})}(1+\frac{1}{\alpha }\sum_{i=1}^{n}\omega
_{i}^{-2}\xi _{i}^{2})^{-\frac{\alpha +n}{2}}\prod_{i=1}^{n}\frac{1}{\omega
_{i}},  \label{St}
\end{equation}
for modeling the empirical PDFs with increments $\xi _{i}$. It is not
difficult to verify that the marginal PDF (\ref{St}) is again PDF (\ref{St}%
). If we integrate out all of the $\xi _{i}$ except for one, we get (\ref{St}%
) with $n=1$. The tails of the distributions behave empirically like \cite
{para} $\sim d\xi /\xi ^{4},$ and so $\alpha \sim 3$.\ In Fig. 1, we compare
the $n=1,$ $\alpha =3$ Student PDF with the Gaussian PDF and with the
empirical univariate PDF constructed for the 100+ years of the daily prices
of the Dow Jones 30 Industrial Average (DJIA),\ starting on the May 26, 1896 
and ending on the December 31, 1999 (i.e. a total of 28507
trading days). The drift rate was determined to be $\mu =0.000257,$ while {\rm Var}%
$[\xi ]=0.000117$. The PDFs are shown as functions of the $\xi /\sigma $
with $\sigma ^{2}={\rm Var}[\xi ]$. The Student PDF fits the empirical data
very well. The common scale parameter is derived from Eq.(\ref{y2}) using
the empirical value of {\rm Var}$[\xi ]$.

For the Student PDF (\ref{St}), we have 
\begin{eqnarray}
{\rm E}[\xi _{i}] &=&0,  \label{x1} \\
{\rm Var}[\xi _{i}] &=&\omega _{i}^{2}\frac{\alpha }{\alpha -2},  \label{y2}
\\
{\rm Corr}[\left| \xi _{i}\right| ^{r},\left| \xi _{k}\right| ^{r}] &=&\frac{%
\Gamma (\frac{r+1}{2})^{2}(\Gamma (\frac{\alpha }{2})\Gamma (\frac{\alpha -2r%
}{2})-\Gamma (\frac{\alpha -r}{2})^{2})}{\sqrt{\pi }\Gamma (\frac{2r+1}{2}%
)\Gamma (\frac{\alpha }{2})\Gamma (\frac{\alpha -2r}{2})-\Gamma (\frac{r+1}{2%
})^{2}\Gamma (\frac{\alpha -r}{2})^{2}}.  \label{x3}
\end{eqnarray}
The correlation of the increments is the same as in the Gaussian random walk
and given by Eq.(\ref{zero}). According to Eqs.(\ref{noise1}), (\ref{noise2}%
), and (\ref{x1}), (\ref{y2}), the square of the volatility equals 
\begin{equation}
\sigma _{i}^{2}=\omega _{i}^{2}\frac{\alpha }{\alpha -2}.  \label{mvol}
\end{equation}
The correlation function (\ref{x3}) is positive and does not depend on the
lag $l=i-j.$ Notice that ${\rm Corr}[\left| \xi _{i}\right| ,\left| \xi
_{k}\right| ]=2/(2+\pi )=\allowbreak 0.\,39$, while the empirical value is
two times lower (see below). The correlation function diverges at $r\geq 
\frac{3}{2}$ due to the heavy tails.

If all components of the vectors ${\bf \psi }=(\psi _{1},...,\psi _{n})$ and 
${\bf \eta }=(\eta _{1},...,\eta _{\alpha })$ are normally distributed $\sim
N(0,1)$, a random vector with components 
\begin{equation}
\xi _{i}=\omega _{i}\psi _{i}{\bf |}\frac{\alpha }{\eta ^{2}}|^{\frac{1}{2}}
\label{rep}
\end{equation}
and $\eta =|{\bf \eta }|$ has the PDF (\ref{St}) (see e.g. \cite{portenko}).
The cumulative effect, $\xi =\sum_{i=1}^{n}\xi _{i},$ is described by 
\begin{equation}
dW(\xi )=\int d\xi \delta (\xi -\sum_{i=1}^{n}\xi _{i})S_{n}^{\alpha }(\xi
_{i})\prod_{i=1}^{n}d\xi _{i}=S_{1}^{\alpha }(\frac{\xi }{\Omega })\frac{%
d\xi }{\Omega }  \label{scaling}
\end{equation}
where $\Omega ^{2}=\sum_{i=1}^{n}\omega _{i}^{2}.\,$The value $\xi $ can be
represented as $\xi =(\omega _{1}\psi _{1}+\omega _{2}\psi _{2}+...+\omega
_{n}\psi _{n}){\bf |}\frac{\alpha }{\eta ^{2}}|^{\frac{1}{2}}\sim \Omega
\psi {\bf |}\frac{\alpha }{\eta ^{2}}|^{\frac{1}{2}}$, where $\psi \sim
N(0,1)$, and so Eq.(\ref{scaling}) easily follows. The variance of the $\xi $
increases linearly with $n,$ in agreement with the empirical observations.
Eq.(\ref{scaling}) represents the exact scaling law for the financial time
series, observed empirically by Mandelbrot \cite{mandelbrot} and discussed
by many authors \cite{para,mantegna,manst}.

The Student conditional PDF at $n\gg l=n-k$ that gives a forecast density $l$
steps ahead has the form 
\begin{equation}
S_{n}^{\alpha }(\xi _{n},...,\xi _{k+1}|\xi _{k},...,\xi _{1})=\frac{%
S_{n}^{\alpha }(\xi _{n},...,\xi _{1})}{S_{k}^{\alpha }(\xi _{k},...,\xi
_{1})}\sim \frac{1}{(2\pi \upsilon ^{2})^{l/2}}\exp (-\frac{1}{2\upsilon ^{2}%
}\sum_{i=k+1}^{n}\xi _{i}^{2})  \label{stcond}
\end{equation}
where 
\begin{equation}
\upsilon ^{2}=\frac{1}{k}\sum_{i=1}^{k}\xi _{i}^{2}.  \label{arch}
\end{equation}
The conditional PDF of the values $\xi _{n},...,\xi _{k+1}\ $is therefore
close to the Gaussian PDF. The ARCH models \cite{engle} propose that the
increments are distributed as $\sim N(0,\upsilon ^{2}),$ with the volatility 
$\upsilon \ $being a function of the lagged increments. The estimator (\ref
{arch}) is one of the possible estimators quantifying the volatility $%
\upsilon \,$within the ARCH framework.

The multivariate Student PDFs have therefore heavy tails and the exact
scaling invariance from the start. The PDFs (\ref{St}) are apparently a
reasonable starting approximation for a precise modeling the empirical PDFs.
These distributions can be modified further to describe two other well
well-established stylized facts which are (iii) long ranged volatility-volatility
correlations that are also known as volatility clustering \cite{ding} and
(iv) return-volatility correlations that are also known as leverage effect 
\cite{black,cox}.

The empirical facts show that there is a slow decay of the correlation
function. An extension of the PDF (\ref{St}) that has the value ${\rm %
Corr}[\left| \xi _{i}\right| ^{r},\left| \xi _{j}\right| ^{r}]$ which is decaying
with time is rather straightforward. From the representation (\ref{rep}) it
is clear that the long ranged correlations occur, since the denominator $%
\sqrt{\eta ^{2}/\alpha }$ is common to all of the increments $\xi _{i}$. In
order to provide a decay of the correlations, it is sufficient to use
different ${\bf \eta }$'s for different groups of the $\psi _{i}$'s. The
analogy with the Ising model can be useful: The components $\psi _{i}$ of
the random vector ${\bf \psi }$ with the same denominator $\sqrt{\eta
^{2}/\alpha }$ can be treated as domains of spins aligned in the same
direction. We assign the usual probability for every such configuration: 
\begin{equation}
w[\sigma _{1},...,\sigma _{n}|\beta ]=N\exp (-\beta \sum_{i=1}^{n-1}\sigma
_{i}\sigma _{i+1})  \label{ising}
\end{equation}
where $\sigma _{i}=\pm 1.$ The normalization constant is given by $%
1/N=2(2\cosh (\beta ))^{n-1}.$ The correlation of the absolute values of the
increments equals (\ref{x3}) provided that $\xi _{i}$ and $\xi _{k}$
belong to the same domain. Otherwise the result is zero. The probability of getting the $%
\xi _{i}$ and $\xi _{k}$ within the same domain can be found to be 
\begin{equation}
w_{l}=e^{-\gamma l}  \label{group}
\end{equation}
where $e^{-\gamma }=e^{-\beta }/(e^{\beta }+e^{-\beta })<1\,$and $l=i-k.$
The coefficient ${\rm Corr}[\left| \xi _{i}\right| ^{r},\left| \xi
_{k}\right| ^{r}]$ for the modified PDF has therefore the form of Eq.(\ref
{x3}) multiplied by $w_{l}.$ To fit the empirical data, we use a
superposition of (\ref{group}) with different values of the $\gamma $. The
absence of the correlations would formally correspond to $\beta =+\infty $ $%
(\gamma =+\infty )$. This is the case when the multivariate PDF is a product
of the $n$ univariate PDFs.

In order to incorporate leverage effect, we consider the PDFs (\ref{St})
with $\omega _{i}\ $depending on the signs $\epsilon _{i-p}={\rm sign}(\xi
_{i-p})$ of the lagged increments ($p=1,2,$ $...$)$.$ The values $\epsilon
_{i}=\pm 1$ are assumed to be independent variables which take the two
values $\pm 1$ with the equal probabilities. The function $\omega _{i}=\chi
(\epsilon _{i-1},\epsilon _{i-2},...)>0$ is defined as 
\begin{equation}
\omega _{i}=C\left( 1-\rho \frac{1-\nu }{\nu }\sum_{p=1}^{\infty }\nu
^{p}\epsilon _{i-p}\right) ^{2}.  \label{omega}
\end{equation}
A requirement of $0<\nu <1$ should be imposed to ensure the convergence of the series. For $%
0\leq \rho <1,$ we have $0<(1-\rho )^{2}\leq \omega _{i}\leq (1+\rho )^{2}$.
Negative recent returns $\epsilon _{i-p}=-1$ ($p=1,2,$ $...$) increase the
volatility, so $\rho >0.$ The normalization condition is taken to be ${\rm E}%
[\omega _{i}]=1,$ so $C=(1+\rho ^{2}\frac{1-\nu }{1+\nu })^{-2}.$ The value
of $\rho $ is connected to the overall strength of leverage effect.

Taking volatility clustering and leverage effect into account, we
obtain: 
\begin{eqnarray}
{\rm Corr}[\left| \xi _{i}\right| ,\left| \xi _{k}\right| ] &=&\frac{2}{\pi }%
\frac{(1-\frac{2}{\pi })g_{l}w_{l}+\frac{2}{\pi }(g_{l}-1)}{g_{0}-\frac{4}{%
\pi ^{2}}},  \label{clus} \\
{\rm Corr}[\left| \xi _{i}\right| ,\xi _{k}] &=&\frac{2}{\pi }\frac{(1-\frac{%
2}{\pi })w_{l}+\frac{2}{\pi }}{\sqrt{(g_{0}-\frac{4}{\pi ^{2}})g_{0}}}h_{l}
\label{leve}
\end{eqnarray}
where $g_{l}={\rm E}[\omega _{i}\omega _{k}]$, $h_{l}={\rm E}[\omega
_{i}\epsilon _{k}\omega _{k}],$ and $l=i-k\geq 0.$ We have 
\begin{eqnarray}
(1+\rho ^{2}\frac{1-\nu }{1+\nu })^{2}g_{l} &=&(1+\rho ^{2}\frac{1-\nu }{%
1+\nu })^{2}+4\rho ^{2}\frac{1-\nu }{1+\nu }\nu ^{l}+4\rho ^{4}\frac{\nu
^{2}(1-\nu )^{3}}{(1+\nu )(1-\nu ^{4})}\nu ^{2l},  \label{gl} \\
(1+\rho ^{2}\frac{1-\nu }{1+\nu })h_{l} &=&-2\rho \frac{1-\nu }{\nu }\nu
^{l}.  \label{hl}
\end{eqnarray}

The empirical correlation functions for the daily prices of the DJIA\ are
fitted using a superposition $w_{l}=c_{1}e^{-\gamma _{1}l}+c_{2}e^{-\gamma
_{2}l}\ $for$\;l>0$ with $c_{1}=0.18,$ $c_{2}=0.08,\,\gamma _{1}=1/1200,$
and $\gamma _{2}=1/233$ (else $c_{0}=0.74\,$and $\gamma _{0}=+\infty ,\,$so
that $\sum_{m=0}^{2}c_{m}=1$)$.$ The values (\ref{ising}) can be interpreted
as conditional probabilities, while the $c_{m}$ are probabilities to get the 
$\gamma _{m}$. We have also used $\rho =1\,$and $\nu =\exp (-1/16).$ The results
shown in Fig. 2 are in a very good agreement with the data. The value ${\rm %
Corr}[\xi _{i},|\xi _{k}|]$ vanishes at $i>k$ in agreement with the
observations, since the $\omega _{i}$ depend on lagged increments only.

Finally, the multivariate PDF of the increments is given by 
\begin{equation}
S_{n}^{\alpha }(\xi _{n},...,\xi
_{1})_{F}=\sum_{m=0}^{2}c_{m}\sum_{s=0}^{n-1}\frac{\exp (-\beta _{m}(n-2s-1))%
}{(2\cosh (\beta _{m}))^{n-1}}\sum_{n_{s},...,n_{0}}%
\prod_{f=1}^{s+1}S_{n_{f}-n_{f-1}}^{\alpha }(\xi _{n_{f}-1},...,\xi
_{n_{f-1}})  \label{final}
\end{equation}
where $n_{s+1}=n+1,$ $n\geq n_{s}>...>n_{1}\geq 2,$ and $n_{0}=1.$ The
number of terms entering Eq.(\ref{final}) and therefore the calculation time 
increases exponentialy with $n$. Using a Pentium IV Microprocessor, the straightforward calculation 
of the $n=13$ PDF (\ref{final}) takes $\sim 10$ sec. The marginal probability of
the PDF (\ref{final})\ is a PDF\ (\ref{final}) again, but for a smaller
sequence of the increments: 
\begin{equation}
\int S_{n}^{\alpha }(\xi _{n},...,\xi _{1})_{F}\prod_{i=k+1}^{n}d\xi
_{i}=S_{k}^{\alpha }(\xi _{k},...,\xi _{1})_{F}.  \label{mar}
\end{equation}

The value of $c_{0}=0.74\,$turned out to be unexpectedly large. It indicates
that approximately $75 \%$ of the empirical PDF consists of the product of the
univariate Student PDFs. The second moments of $\xi _{i}$ are finite, so the
product does converge to the Gaussian PDF due to the central limit theorem 
\cite{portenko}. One should thus verify that the PDF (\ref{final})
nevertheless supports the approximate scaling observed for financial time
series \cite{mandelbrot,para,manst}. In Fig. 3, we plot the scaled PDFs of
the values $\xi =\sum_{i=1}^{n}\xi _{i}$ for $n=1,2,3,5,8,$ and $13$ as
functions of $\xi /{\rm Var}[\xi ]^{\frac{1}{2}},$ together with the scaled
Gaussian PDF. We see that the convergence to the Gaussian PDF is very slow.
A qualitative explanation of this phenomenon is based on the fact that the
L\'{e}vy stable PDF fits the central part of the empirical distributions
\cite{para,mantegna} and therefore the central part of the Student PDF (see Fig. 1),
providing an exceptionally slow convergence to the Gaussian PDF. In Fig. 4,
we compare the scaled $n=13\,$distributions based upon the PDF (\ref{final})
with the empirical $n=13$ PDF. The results are in excellent agreement,
whereas the $n=1$ Student PDF and the Gaussian PDF disagree clearly with the
data.

The PDF (\ref{final}) is constructed for the daily increments. There are no 
{\it a priori} reasons to believe that the daily scale is more appropriate
than other time scales. The criterion for selection of the best time
interval is the quality of the fit of the empirical univariate PDF. It is
seen from Fig. 4 that e.g. $\Delta t=13\,$days would be an inaccurate
choice. At the same time, the excellent fit of the daily empirical PDF,
shown in Fig. 1, indicates that the assumption $\Delta t=1$ day on which the
PDF (\ref{final}) is based is quite reasonable.

If we work with an ensemble of uncorrelated stocks and calculate ensemble
averages, we should use the marginal probabilities. If we work with a
single index, we only calculate the time averages. In such a case, the
conditional probabilities should be used. The previous discussion refers to the
calculation of the marginal probabilities. If no correlations exist, as in
the case of the Gaussian random walk, the marginal probabilities coincide
with the conditional probabilities.

We now present some results for the time averages represented by the conditional probabilities. 
They are distinct from the marginal ones, since
the absolute values of the increments $\xi _{i}$ are correlated. In our
case, the ergodic hypothesis is valid, since all of the values $\gamma _{k}$
and ${\rm ln}(1/\nu )$ are distinct from zero, so the correlation lengths
are all finite. This means that the time averages are equivalent to the
ensemble averages, provided that the time scale used for the estimation is
much longer than the largest correlation length ($1/\gamma _{1}\sim 5$
calendar years). Using smaller time scales, however, the distinction 
between these two types of the probabilities can be significant.  

The conditional PDF of the increments $\xi _{n},...,\xi _{k+1}|\xi
_{k},...,\xi _{1}$ equals 
\begin{equation}
S_{n}^{\alpha }(\xi _{n},...,\xi _{k+1}|\xi _{k},...,\xi _{1})_{F}=\frac{%
S_{n}^{\alpha }(\xi _{n},...,\xi _{1})_{F}}{S_{k}^{\alpha }(\xi _{k},...,\xi
_{1})_{F}}.  \label{copr}
\end{equation}
It is normalized to unity according to Eq.(\ref{mar}). Note that for $i>k$%
\begin{eqnarray}
{\rm E}[\xi _{i}|\xi _{k},...,\xi _{1}]_{F} &=&0,  \label{av} \\
{\rm Var}[\xi _{i}|\xi _{k},...,\xi _{1}]_{F} &=&2^{-(i-k)}\sum_{\epsilon
_{i},...,\epsilon _{k+1}}\sum_{m=0}^{2}c_{m}\sum_{v=1}^{i}\frac{\exp (-\beta
_{m}(i-v-\varkappa ))}{(2\cosh (\beta _{m}))^{i-v+\varkappa }}  \nonumber \\
&&\times \omega _{i}^{2}\frac{\alpha }{\alpha -2+\varkappa _{1}}(1+\frac{1}{%
\alpha }\sum_{p=v}^{k}\omega _{p}^{-2}\xi _{p}^{2})  \label{va}
\end{eqnarray}
where $\varkappa =\min (1,v-1),$ and $\varkappa _{1}=\max (0,k+1-v).$ The
conditional volatility, according to Eqs.(\ref{noise1}) and (\ref{noise2}),
equals 
\begin{equation}
\sigma _{ii-1}^{2}={\rm Var}[\xi _{i}|\xi _{i-1},...,\xi _{1}]_{F},
\label{cvol}
\end{equation}
while the volatility (\ref{mvol}) can be referred to as the marginal
(unconditional) volatility. The ensemble average of (\ref{cvol}) provides (\ref
{mvol}).

The value $\Delta _{i}\equiv \ln S_{i}/S_{i-1}\,$determines the log-return on an asset
per interval $[t_{i},t_{i-1}].$ The increments $\xi _{i}\sim 10^{-2}$ and
the drift rate $\mu \,$are small, and so $\Delta _{i}\simeq \xi _{i}$ to a
first approximation. The Ito's lemma (see e.g. \cite{hull}) can be applied
to $\ln S(t)$ where $S(t)$ is governed$\;$by Eq.(\ref{dS/S}) provided that $%
{\rm E}[(d\xi (t))^{n}]<\infty $ for all $n$. In our case, the asymptotic
distributions of the increments are such that$\ {\rm E}[(d\xi
(t))^{2n}]=\infty $ for $n\geq 2$. Eq.(\ref{dS/S}) is therefore equivalent
to equation
\begin{equation}
\Delta _{i}=\ln (1+\mu +\xi _{i})  \label{dlnS}
\end{equation}
that cannot be simplified further by expanding the logarithm in the power series
of the $\xi _{i}\,$and discarding the averages of the higher order terms in $%
\xi _{i}$.

The multivariate PDF of log-returns can be written as $dW[\Delta
_{i}]=S_{n}^{\alpha }(\xi _{i})_{F}\prod_{i=1}^{n}d\xi _{i}$ where $\xi _{i}$
is defined directly by Eq.(\ref{dlnS})$.$ The knowledge of the multivariate
PDF provides, in principle, the most complete information on the future
market behavior.

To make a simple check, let us consider the fair value of a stock or a
market index at $t=t_{k}$: 
\[
{\rm E}[S_{n}]=S_{k}{\rm E}[\exp (\sum_{i=k+1}^{n}\Delta _{i})|\xi
_{k},...,\xi _{1}]_{F}=(1+r)^{l}S_{k} 
\]
where $\mu =r$ is the risk-free discount rate corresponding to the
expiration date $t_{n}$. The result has the conventional form. The
dispersion of $S_{n}$, however, diverges: {\rm Var}$[S_{n}|\xi _{k},...,\xi
_{1}]_{F}=\infty .$ One can probably assume the existence of a cutoff in the
PDFs at $|\xi _{i}|\sim \Lambda $. The largest one-day
variation of the DJIA took place on the October 19,  1987, so the value of $%
\Lambda $ is constrained from below by $\Lambda $ $\gtrsim 0.25.$

To illustrate possible applications of the model, let us consider pricing of
the log contracts \cite{Log}. The fair value of a long position, $L=\ln
S_{n}/S_{k},$ is defined as the conditional expectation value: 
\[
{\rm E}[L]={\rm E}[\sum_{i=k+1}^{n}\Delta _{i}|\xi _{k},...,\xi _{1}]_{F}. 
\]
The integrals are determined by the region $|\xi _{i}|\sim 10^{-2}.$ The
values $\Delta _{i}$ can be expanded in the power series up to $O(\xi _{i}^{2}),$
in which case the integrals are well defined. Using the conditional PDF (\ref
{copr}), we obtain: 
\begin{equation}
{\rm E}[\sum_{i=k+1}^{n}\Delta _{i}|\xi _{k},...,\xi _{1}]_{F}\simeq
2^{-l}\sum_{\epsilon _{n},...,\epsilon _{k+1}}\sum_{i=k+1}^{n}(r-\frac{1}{2}%
\sigma _{ik}^{2}).  \label{fv}
\end{equation}
The result depends on the lagged increments through $\sigma _{ik}^{2}={\rm %
Var}[\xi _{i}|\xi _{k},...,\xi _{1}]_{F}$ given by Eq.(\ref{va}). The
pricing of log contracts is well defined both with respect to the fair
value and the dispersion.

It is hard to distinguish empirically between the increments $\xi _{i}$ and
log-returns $\Delta _{i}$. The small distinction results in particular to small
return-return correlations due to the non-vanishing correlations of $\sigma
_{i}^{2}$ and $\xi _{k}.$ This effect is, however, statistically not
significant for empirical tests. The asymptotic $dW\sim d\xi /\xi ^{4}\sim
e^{-3\Delta }d\Delta $ implies {\rm E}$[S_{n}]<\infty ,\,$and {\rm Var}$%
[S_{n}]<\infty ,\,$while $dW\sim d\Delta /\Delta ^{4}\sim d\xi /(\xi \ln
^{4}\xi )$ would cause severe theoretical problems: {\rm E}$[S_{n}]=$ 
{\rm Var}$[S_{n}]=\infty .$ The asymptotic behavior $\sim
d\xi /\xi ^{4}$ refers apparently to the increments $\xi _{i}$.

The asymptotic behavior of the return distribution is important for correct
estimates of financial risks. The Gaussian walk underestimates the
probability of large fluctuations, and in particularly, the probability of
crushes. The use of the multivariate Student PDF (\ref{final}) gives a more
precise idea on the involved risks, and in particular, on those
connected to large market fluctuations.

The following four stylized facts (i) heavy tails, (ii) scaling of the
returns, (iii) volatility clustering and (iv) leverage effect are
well established empirically, but missing in the Gaussian random walk model. We
constructed the multivariate PDFs of incremenets and returns for financial
markets, that take approximately those four effects into account. This model
can be useful for more accurately pricing derivative securities and risk
management.

\vspace{1cm}

The authors wish to thank the Dow Jones Global Indexes for providing 
the DJIA historical prices and the permission to
use results based on the analysis of these prices for present publication.
M.I.K. is grateful to Metronome-Ricerca sui Mercati Finanziari for kind
hospitality.

\newpage{}

\begin{figure}[h]
\vspace{2cm}
\begin{center}
\leavevmode
\epsfxsize = 12cm
\epsffile[20 40 573 465]{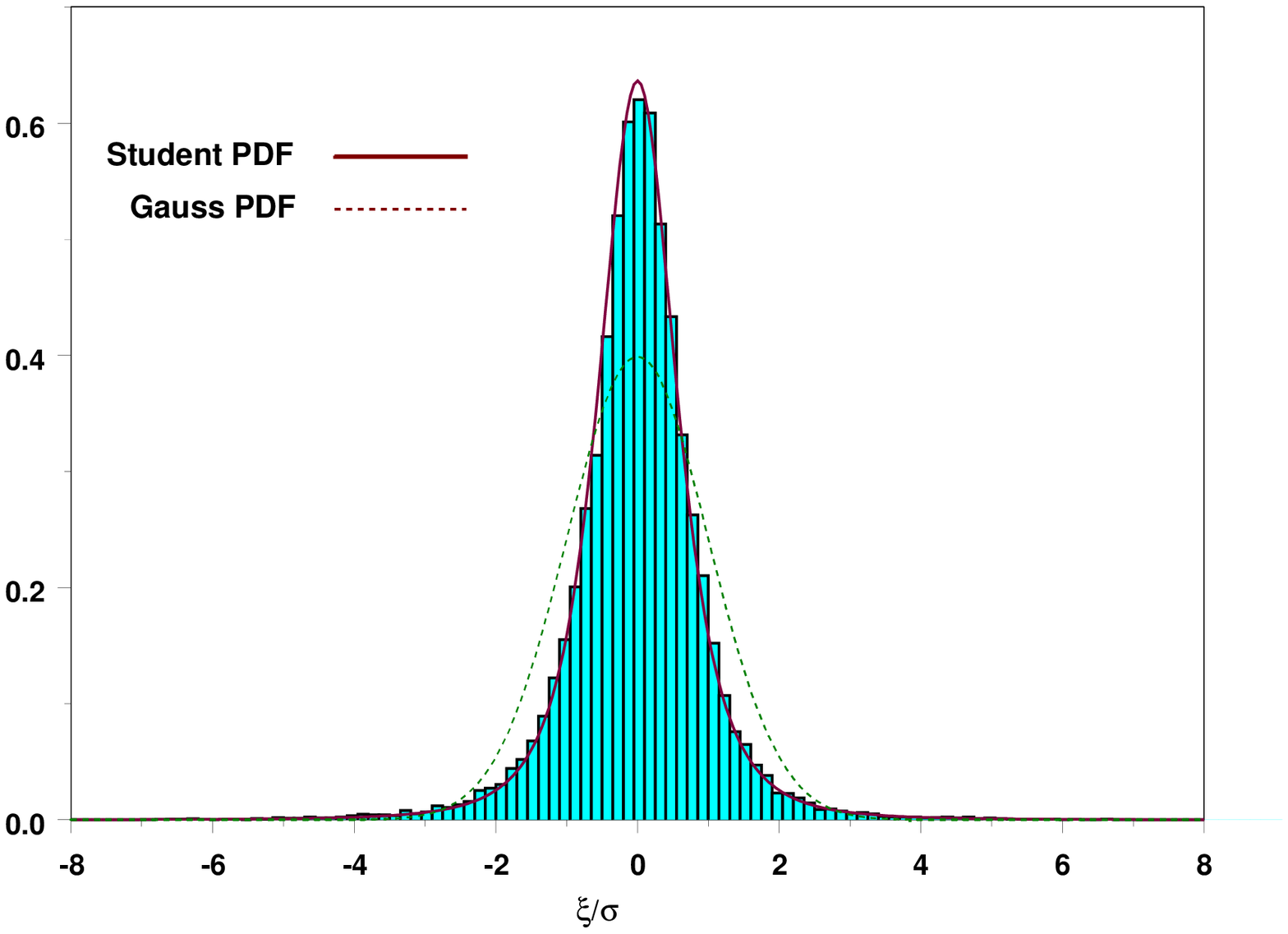}
\end{center}
\caption{
The $\alpha = 3$ Student and Gaussian univariate PDFs as compared to the hystogram for the 100+ years of daily prices
of the Dow Jones 30 Industrial Aveage. The value of $\xi$  is the increment (noice) added to the path followed by the index value,
$\sigma ^{2}={\rm Var}[\xi ]$. The common scale of the both distributions is fixed by fitting the variance. The solid line describes the Student
PDF, the short-dashed line corresponds to the Gaussian PDF.
}
\label{fig1}
\end{figure}

\begin{figure}[h]
\vspace{2cm}
\begin{center}
\leavevmode
\epsfxsize = 12cm
\epsffile[23 40 573 465]{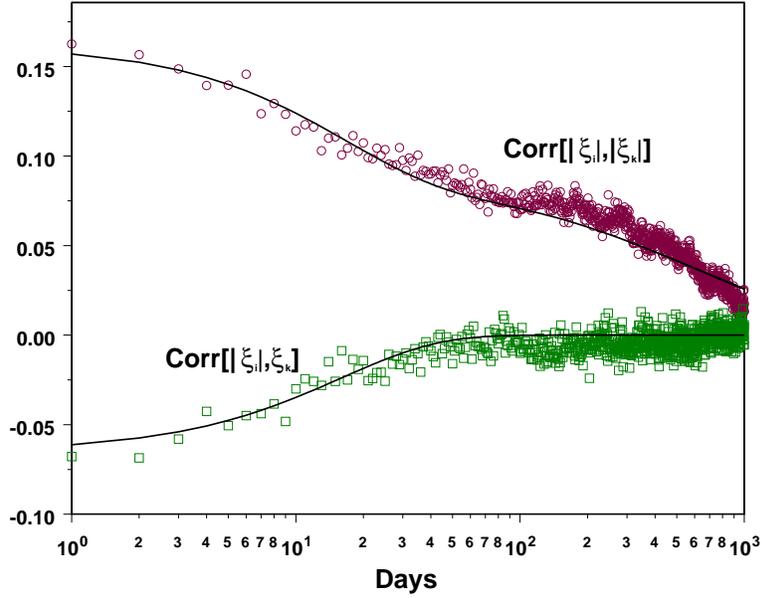}
\end{center}
\caption{
Correlation functions ${\rm Corr} [|\xi_{i}|,|\xi_{k}|]$ and
${\rm Corr} [|\xi_{i}|,\xi_{k}]$  
versus time lag $l=i-k$ between two increments $\xi_{i}$ and $\xi_{k}$. 
The empirical data are constructed using the
daily prices of the DJIA. The volatility clustering (circles) 
and the leverage effect 
(boxes) can be clearly seen. The data is compared to predictions of the multivariate Student PDF, 
shown by the solid lines.
}
\label{fig2}
\end{figure}

\begin{figure}[h]
\vspace{2cm}
\begin{center}
\leavevmode
\epsfxsize = 12cm
\epsffile[23 40 573 465]{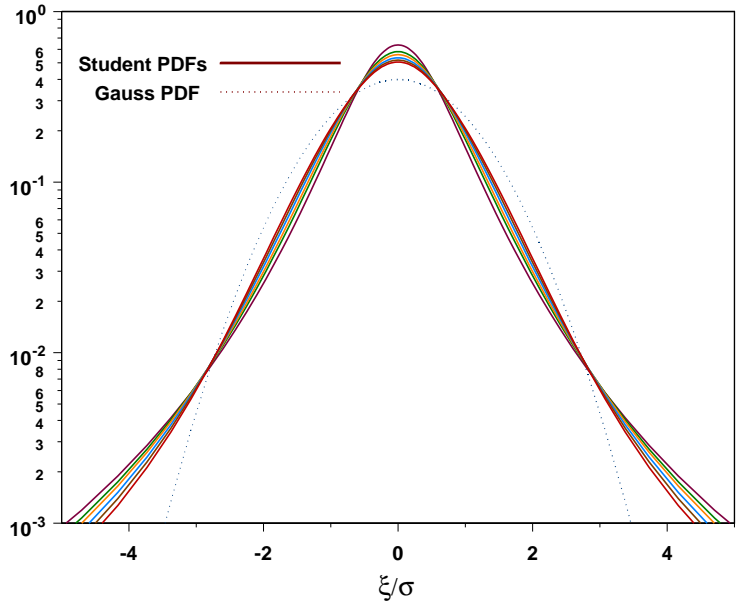}
\end{center}
\caption{
The $\alpha = 3$ scaled Student PDF for $n = 1,2,3,5,8,$ and $13$ (solid lines) 
as compared to the Gaussian PDF (short-dashed line). The lower vales of the
$n$ correspond to the higher values of the PDF at the origin $\xi = 0$ and 
at high absolute values of the $\xi$. The convergence to the 
Gaussian PDF with increasing the $n$ is slow. 
The value of $\xi$ is the noise added to the path followed by the index and 
$\sigma^2$ is variance of the $\xi$.
}
\label{fig3}
\end{figure}

\begin{figure}[h]
\vspace{2cm}
\begin{center}
\leavevmode
\epsfxsize = 12cm
\epsffile[23 40 573 465]{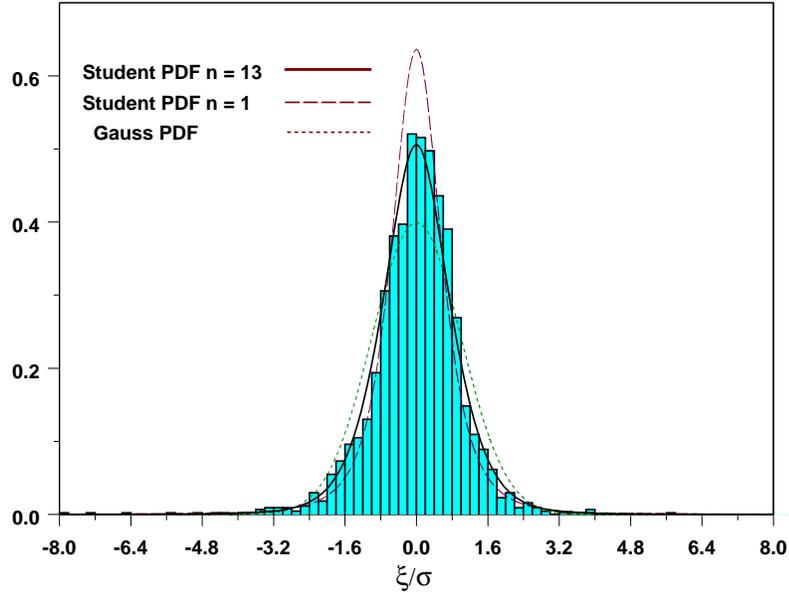}
\end{center}
\caption{
The $\alpha = 3$ $n = 13$ and $n = 1$ Student and Gaussian PDFs as compared to the hystogram for 
the 100+ years of prices
of the DJIA with sampling  $\Delta t = 13$ days. 
The value of $\xi$  is the increment (noice) added to the path followed by the index value,
$\sigma ^{2}={\rm Var}[\xi ]$. The common scale of the distributions is fixed by 
fitting the variance 
of the $\xi$. The solid line stands for the $n = 13$ Student PDF, the long-dashed line stands 
for the $n = 1$ Student PDF, and the short-dashed line
denotes the Gaussian PDF.
}
\label{fig4}
\end{figure}

\end{document}